\documentclass[a4paper]{article}
\usepackage{icrc2013}
\usepackage[english]{babel}
\title{Pointing Calibration for the Cherenkov Telescope Array Medium Size Telescope Prototype
}

\authors{
 L. Oakes$^1$, B. Behera$^2$, J. Baehr$^2$, S. Gruenewald$^1$, T. Raeck$^2$, S. Schlenstedt$^2$, A. Schubert$^2$, U. Schwanke$^1$ \\for the CTA Consortium
}

\afiliations{
$^1$ Humboldt Universit\"at zu Berlin \\
$^2$ DESY, Zeuthen \\
}

\email{loakes@physik.hu-berlin.de}

\abstract{Pointing calibration is an offline correction applied in order to obtain the true pointing
direction of a telescope. The Cherenkov Telescope Array (CTA) aims to have the precision to determine the position of point-like
as well as slightly extended sources, with the goal of systematic errors less than 7 arc seconds
in space angle. This poster describes the pointing calibration concept being developed for the
CTA Medium Size Telescope (MST) prototype at Berlin-Adlershof, showing test results and preliminary measurements.
The MST pointing calibration method uses two CCD cameras, mounted on the telescope dish,
to determine the true pointing of the telescope.  The ``Lid CCD'' is aligned to the optical axis
of the telescope, calibrated with LEDs on the dummy gamma-camera lid; the ``Sky CCD'' is pre-aligned
to the Lid CCD and the transformation between the Sky and Lid CCD camera fields of view is precisely
modelled with images from special pointing runs which are also used to determine the pointing model. 
During source tracking, the CCD cameras record images which are analysed offline
using software tools including Astrometry.net to determine the true pointing coordinates.
}

\begin{document}
\maketitle

\section{Introduction}
Telescope pointing calibration plays an important role in acquiring valid physics measurements, by ensuring the true direction of observations is known. This requires the conversion of telescope drive system coordinates to sky coordinates, and corrections due to local and instrumental conditions. In most cases, this is an offline calibration applied to the mechanical pointing position to obtain the true pointing direction of the observations which have been carried out. 

The Cherenkov Telescope Array (CTA) is a future observatory designed to surpass all current cherenkov telescope experiments by up to an order of magnitude in sensitivity and precision. The intended pointing precision for CTA is to be able to determine the position of point-like and slightly extended sources with a systematic uncertainty of less than 7 arcsec. CTA will be made up of 3 sizes of telescope, Small Size Telescopes (SSTs) of 4-6m diameter sensitive in the high energy ($>10$ TeV range,  Medium Size Telescopes (MSTs) of 10-12m diameter for the core energy range of 100 GeV to a few TeV and Large Size Telescopes (LSTs) with a diameter of 23m, sensitive to the lowest energy gammas. 

A prototype MST is in the final stages of construction in Berlin-Adlershof, this poster describes the development of a pointing calibration technique for the MST prototype, which will be used to inform decisions about pointing calibration for the final set of CTA MSTs. 


\section{The MST Prototype at Adlershof}
\begin{figure}[ht]
  \centering
  \includegraphics[width=0.4\textwidth]{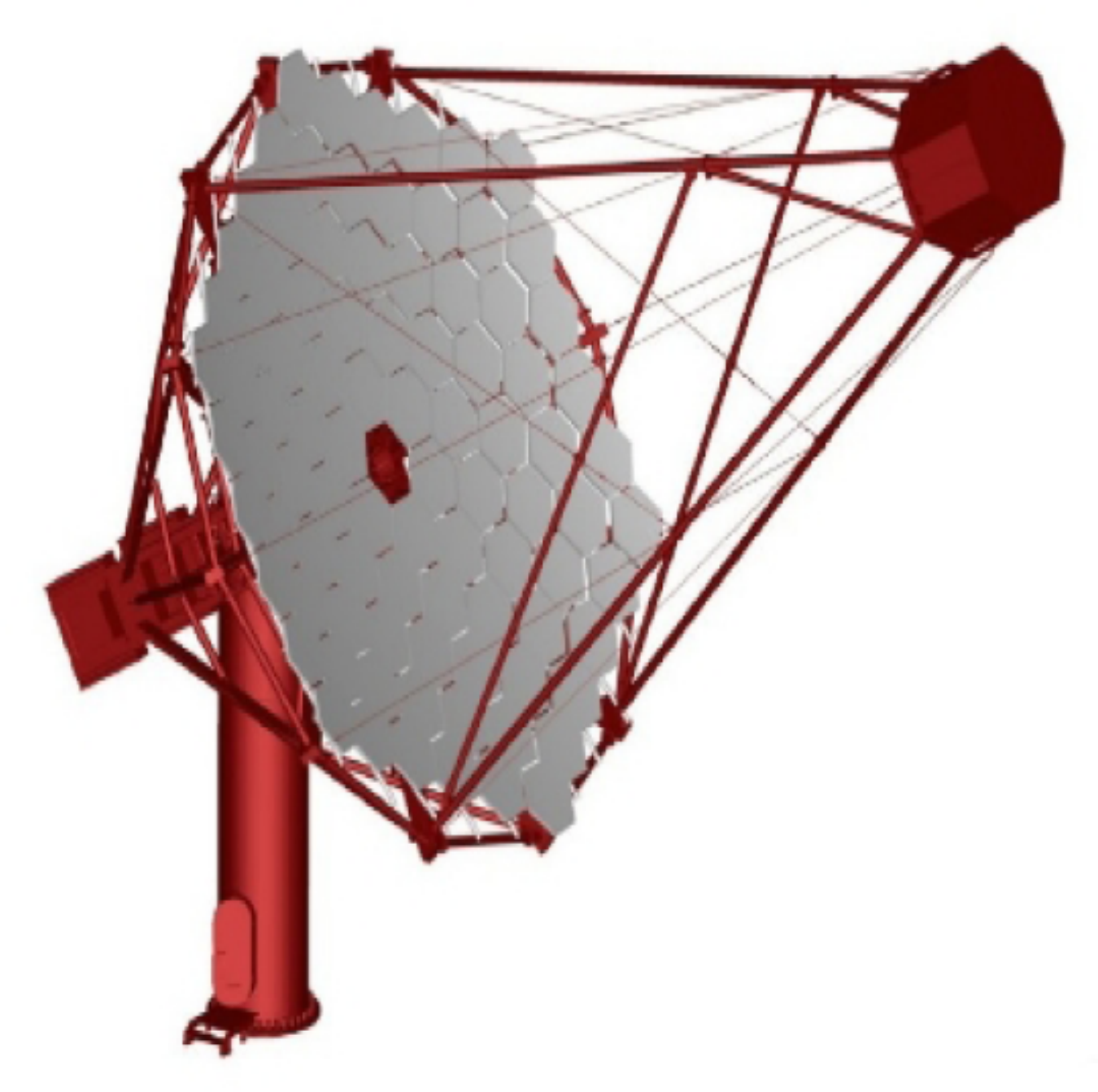}
  \caption{Design of the MST prototype}
  \label{fig1}
 \end{figure}

The prototype telescope under construction in Berlin-Adlershof is a full mechanical prototype of the modified Davies-Cotton type MST for CTA~\cite{bib:proto}. It comprises a full drive system, a 12m diameter optical support structure (``dish'') with a mixture real and dummy mirror segments, active mirror control (AMC) and a dummy camera with realistic weighting. The 84 mirror facets mounted on the dish are hexagonal, spherical mirrors, which are adjustable in 3D space. The prototype is controlled by array control software (ACTL) based on the ACS (ALMA Common Software) framework~\cite{bib:actl}, which is being developed for wider CTA use. The ACTL software is used to control and read out devices including the drive system, CCD cameras for pointing and optical calibration, a weather station and sensors; this provides an ideal testing ground for the software.  Figure~\ref{fig1} shows the design concept of the MST prototype. 

While star visibility in Berlin is not optimal for physics observations, it has been shown that the sky conditions are sufficient for accurate pointing measurements. Tests of the telescope pointing and pointing calibration are an important part of the prototype programme, along with measurement of the Point Spread Function (PSF), safety system tests, observation of weather conditions and related performance, testing of ACTL and proof of the mechanical concept.

\section{Pointing Calibration for the MST prototype}
The following subsections give details of the hardware and software used to calibrate the MST prototype pointing.
\subsection{Instrumentation}
Three CCD cameras, housed in weatherproof casing are installed on the prototype dish, of these two are used for pointing calibration. The central ``LidCCD'' camera is mounted on the central plate, pointing at the dummy camera. LEDs on the dummy camera lid allow determination of the camera position with respect to the LidCCD. This camera is also used for mirror alignment and measurement of the PSF.

The second camera (``SkyCCD'') is mounted on the edge of the dish (6m from the centre) to take unobstructed images of the sky in the direction of the telescope pointing. This camera must be precisely aligned to the LidCCD, as described in the following subsection.

The CCD cameras are of the type Prosilica GC 1350, with a chip size of 1360x1024 pixels, where a pixel is 4.65x4.65 $\mu$m$^2$. For a large light collection area combined with a suitable field of view, the SkyCCD uses an 85mm Walimex Pro lens with an aperture of f/1.4 . The large lens weighs about 480g and requires a support structure within the camera housing to avoid bending or distortion of the CCD camera. The resulting field of view is 4.26$^{\circ}$ x 3.21$^{\circ}$, which is sufficient for astrometry measurements as well as single bright star observations. 
 
Sensors installed on the CCD cameras record temperature information, to observe the effects of this variable such as CCD chip expansion due to heat, and calibrate where necessary.
 \begin{figure}[ht]
  \centering
  \includegraphics[width=0.4\textwidth]{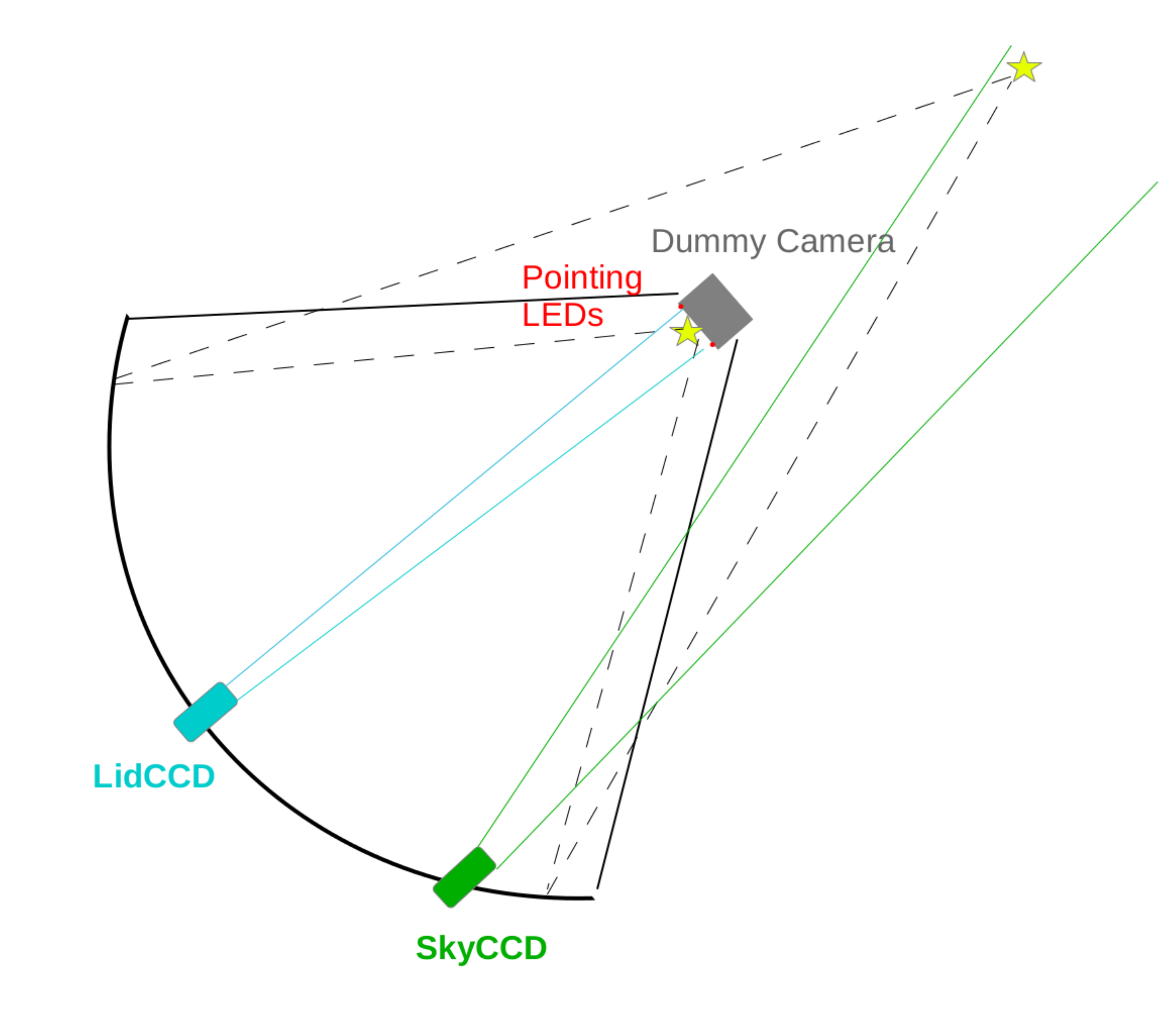}
  \caption{Sketch of 2 CCD camera pointing setup}
  \label{fig2}
 \end{figure}
\subsection{Technique}
The 2 CCD camera pointing technique used for the MST prototype is based on the H.E.S.S. experiment precision pointing method~\cite{bib:hesspp}. 

After alignment of the LidCCD to the telescope optical axis, initial approximate alignment of the two CCDs is carried out in the absence of the dummy camera. The pointing direction of both cameras is adjusted to observe a local object (at a distance of around 1000m), such that an object at infinity in the centre of the field of view of the LidCCD (or telescope optical axis) will be within the field of view of the SkyCCD. This preliminary alignment does not require alignment of the mirrors. 

With the CCD cameras thus aligned, a pointing run can be carried out after mirror alignment, with the dummy camera in place. In this mode, the dummy camera lid is closed, and the image of a section of the sky is focused onto the lid by the mirrors.  Mapping the reflected image to that seen by the SkyCCD gives a transformation between the LidCCD and SkyCCD fields of view. The relevant transformation is calculated for a full range of pointing directions, to account for bending of the telescope frame due to the dummy camera weight, which is dependent on the altitude and azimuth directions. This set of transformations will be stored in a database and can then be used offline to calculate and correct the true pointing direction of the telescope optical axis during normal data taking runs, where the SkyCCD records sky images for calibration. The pointing direction from SkyCCD images is converted to astronomical coordinates using Astrometry.net software.
 
\subsection{Astrometry.net}
The Astrometry.net package~\cite{bib:astrom} takes a sky image with any depth of field, orientation and field of view, and returns the pointing, scale and orientation of the image.  The software identifies sets of 4-5 stars in the image, and compares these with pre-indexed images. These indices are built from the USNO-B and 2MASS star catalogues, and the package has a 100$\%$ success rate of image identification (assuming sufficient image quality). Quality of multi-star images using the SkyCCD camera and selected optics have been tested as input for Astrometry.net prior to installation on the prototype. Astrometry.net is already in use for pointing calibration by the VERITAS collaboration~\cite{bib:veritas}.

\subsection{Preliminary studies}
\begin{figure}[ht]
  \centering
  \includegraphics[width=0.4\textwidth]{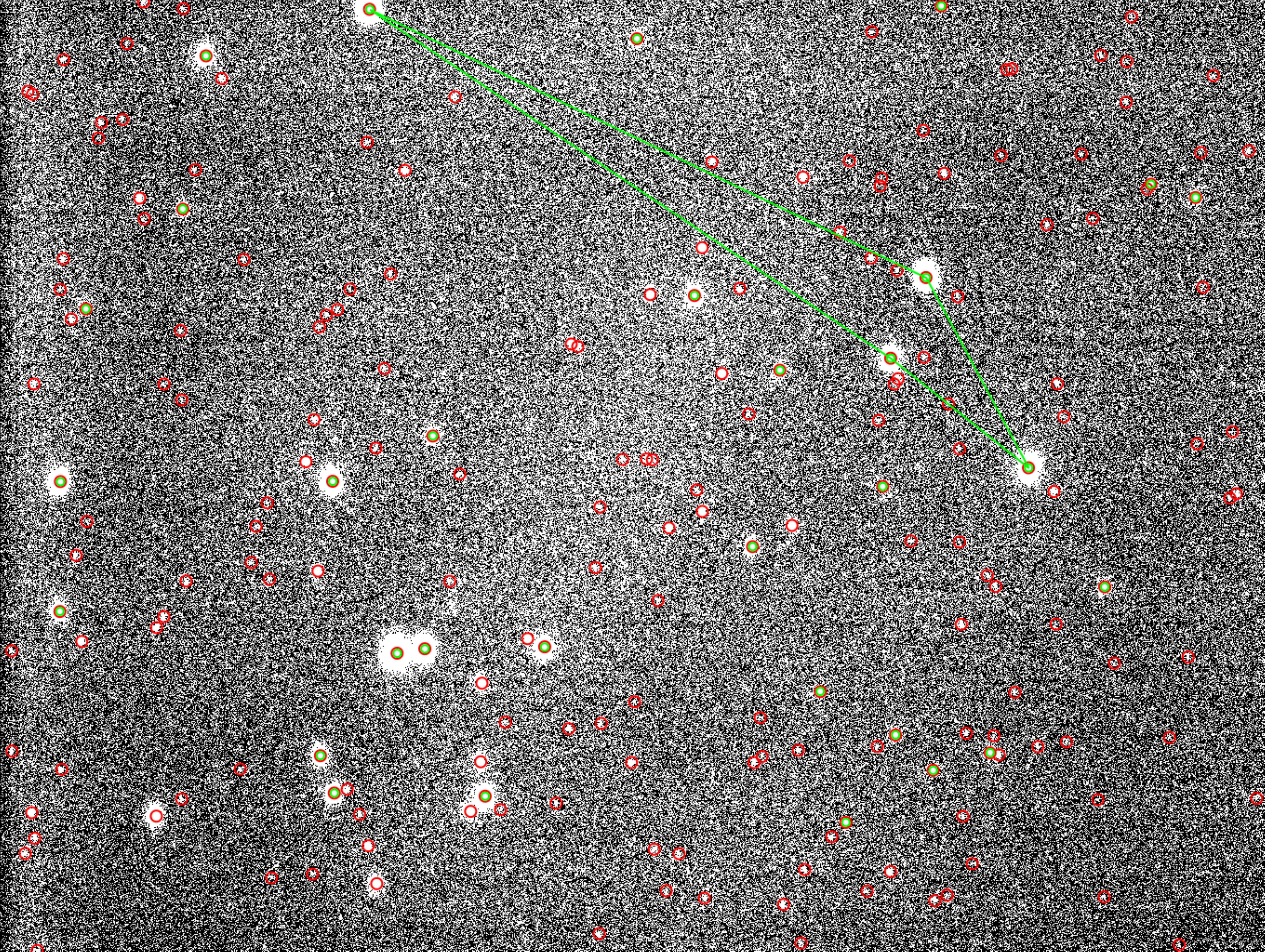}
  \caption{Solved image taken with SkyCCD camera, processed with Astrometry.net software. Green lines and red points indicate the stars identified and indexed by Astrometry.net}
  \label{fig3}
 \end{figure}

Early proof of principle tests have been carried out with the SkyCCD camera and optical equipment which will be used on the MST prototype. Test images were taken using a tracking tripod for a range of gains and exposures to examine the CCD camera performance and lens suitability under various conditions. Figure~\ref{fig3} shows an image with a 3 second exposure, overlaid with indexing information from Astrometry.net. The pointing direction returned for this image is (04:25:58.819, +16:36:45.569) in RA-Dec coordinates.

The successful calculation of the pointing direction from this image suggests that this technique requiring multiple visible stars can be applied despite disfavrouable background light conditions in the region.

\subsection{Systematic uncertainties}
In order to understand the accuracy of the pointing calibration, systematic uncertainties on the calculated pointing direction must be assessed. For this purpose, a single-bright-star image analysis method is being developed to find the pointing direction independently of the astrometry technique. A comparison of the calculated directions from the two methods will indicate the level of systematic uncertainty to be expected. This comparison requires a field of view with the presence of a known bright star, as well as enough other stars to perform astrometry.  Systematic uncertainties will also be assessed within the Astrometry.net framework. 

\section{Conclusions and future}
Pointing calibration for CTA is being developed and tested on the Medium Size Telescope (MST) prototype telescope in Berlin-Adlershof. It was demonstrated by preliminary (pre-prototype installation) measurements, that the sky conditions in this area are suitable for obtaining precision pointing measurements. Understanding gained from pointing measurements with the MST prototype will feedback into the final MST design, including choice of hardware and software development. The MST prototype is in the final construction stages, and first measurements from the installed instruments will be available by summer 2013. MST production will begin in 2015 for installation on site in 2016.

%
%
%
%
%
%
%
%
%
%
%
%
%
%
%
%
%

%
\vspace*{0.5cm}
\footnotesize{{\bf Acknowledgment:}{We gratefully acknowledge support from the agencies and organisations listed at the following URL: http://www.cta-observatory.org/?q=node/22}}


\begin{thebibliography}{}
\bibitem{bib:proto}  Davis, E. et al 2011, Mechanical design of a medium-size telescope for CTA {\it CTA Internal Note}, B\"ahr, J. for the CTA Consortium, Status of the CTA medium size telescope prototype, AIP Conf. Proc. 1505 (2012) 
\bibitem{bib:actl} Oya, I. et al 2012, Evaluating the control software for CTA in a medium size telescope prototype, JoP: Conference Series 396 (2012) 012037 
\bibitem{bib:hesspp} Braun, I. 2007, Improving the Pointing Precision of the H.E.S.S. Experiment, {\it Dissertation, University of Heidelberg} 
\bibitem{bib:astrom} {Lang}, D. et al, Astrometry.net: Blind astrometric calibration of arbitrary astronomical images,  The Astronomical Journal 139, 1782-1800
\bibitem{bib:veritas} VERITAS website: http://veritas.sao.arizona.edu/
\end{thebibliography}
\end{document}